\documentclass[12pt,two column]{iopart}
\usepackage{graphicx}
\usepackage{amssymb}
\begin{document}

\title[Upper critical field, Hall effect and magnetoresistance in LaFeAsO$_{0.9}$F$_{0.1-\delta}$]
{Upper critical field, Hall effect and magnetoresistance in the iron-based layered superconductor LaFeAsO$_{0.9}$F$_{0.1-\delta}$}

\author{Xiyu Zhu, Huan Yang, Lei Fang, Gang Mu and Hai-Hu Wen}

\address{National Laboratory for Superconductivity, Institute of Physics and\\
Beijing National Laboratory for Condensed Matter Physics, P. O. Box
603,\\
Beijing, 100190, P. R. China} \ead{hhwen@aphy.iphy.ac.cn (Hai-Hu
Wen)}

\pacs{74.70.-b, 74.25.Fy, 73.43.Qt}

\begin{abstract}
By using a two-step method, we successfully synthesized the iron
based new superconductor LaFeAsO$_{0.9}$F$_{0.1-\delta}$. The
resistive transition curves under different magnetic fields were
measured, leading to the determination of the upper critical field
$H_{\mathrm{c}2}(T)$ of this new superconductor. The value of
$H_{\mathrm{c}2}$ at zero temperature is estimated to be about 50
Tesla roughly. In addition, the Hall effect and magnetoresistance
were measured in wide temperature region. A negative Hall
coefficient $R_\mathrm{H}$ has been found, implying a dominant
conduction mainly by electron-like charge carriers in this material.
The charge carrier density determined at 100$\;$K is about $9.8
\times10^{20}\;\mathrm{cm}^{-3}$, which is close to the cuprate
superconductors. It is further found that the magnetoresistance does
not follow Kohler's law. Meanwhile, the different temperature
dependence behaviors of resistivity, Hall coefficient, and
magnetoresistance have anomalous properties at about 230$\;$K, which
may be induced by some exotic scattering mechanism.
\end{abstract}

\section{Introduction}
The recently
discovered\cite{FirstJACS,HHwen,xhchen,gfchen,zhianren,zhanren,chengpeng}
iron-based superconducting oxides with a transition temperature as
high as 55K has attracted much attention. As long as a new
superconductor is discovered, the superconducting mechanism is
urgent to be understood and the transition temperature is hoped to
be enhanced. Thus the fundamental parameters, such as the upper
critical field $H_{\mathrm{c}2}$, charge carrier density and its
type, the electron scattering mechanism, etc, are very important to
identify the superconducting mechanism. It is well perceived that
the normal state properties, such as the Hall effect and
magnetoresistance, are very important to learn about the electronic
scattering feature which is intimately related to the mechanism of a
superconductor. For example, the superlinear temperature dependence
of the normal state resistivity and the clear temperature dependence
of Hall coefficient in wide temperature region in cuprate
superconductors have been considered as two important anomalous
properties which suggest the unusual scattering process beyond the
electron-phonon scattering. In this paper we report the successful
fabrication of this material with a two-step method, and the
measurements on the resistive transition under different magnetic
fields, Hall effect and magnetoresistance. By analyzing the data we
obtained fresh information about the upper critical field, the
charge carrier density and its type, as well as the scattering times
etc.

\section{Sample preparation and characterization}
\begin{figure}
\includegraphics[width=8cm]{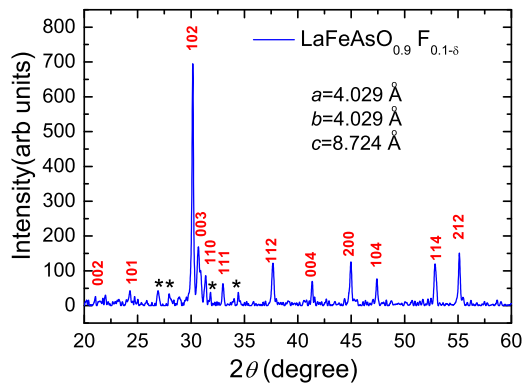}
\caption{(Color online) X-ray diffraction patterns for the sample
LaFeAsO$_{0.9}$F$_{0.1-\delta}$. Almost all main peaks can be
indexed by a tetragonal structure with $a=b=4.029\mathrm{\AA}$ and
$c=8.724\mathrm{\AA}$. The tiny peaks marked by asterisks are from
the impurity phase, perhaps FeAs.} \label{fig1}
\end{figure}

The polycrystalline samples were synthesized by using a two-step
solid state reaction method. First the starting materials Fe powder
(purity 99.95\%) and As grains (purity 99.99\%) were mixed in 1:1
ratio, ground and pressed into a pellet shape. Then it was sealed in
an evacuated quartz tube and followed by heat treating  at
700$^\circ$C for 10 hours. The resultant pellet was smashed and
ground together with the LaF$_3$ powder (purity 99.95\%),
La$_2$O$_3$ powder (purity 99.9\%) and grains of La (purity 99.99\%)
in stoichiometry as the formula LaFeAsO$_{0.9}$F$_{0.1-\delta}$.
Again it was pressed into a pellet and sealed in an evacuated quartz
tube and burned at about 940$^\circ$C for 2 hours, followed by a
heat treating  at 1150$^\circ$C for 48 hours. Then it was cooled
down slowly to room temperature. Since a little amount of F may
escape during the second step fabrication, in the formula for our
sample, we use $0.1-\delta$ as the possible concentration of F. In
Figure.~\ref{fig1}, we show the X-ray diffraction patterns for the
sample LaFeAsO$_{0.9}$F$_{0.1-\delta}$. Almost all main peaks can be
indexed by a tetragonal structure with $a=b=4.029\mathrm{\AA}$ and
$c=8.724\mathrm{\AA}$. Therefore the dominant component is from
LaFeAsO$_{0.9}$F$_{0.1-\delta}$ with only tiny peaks in the same
scale as appeared in the data of the original paper\cite{FirstJACS}.
This tiny amount of second phase, perhaps from FeAs, together with
the granular behavior of the samples, may give some influence on the
zero resistance point, but they should not give any obvious
influence on the upper critical field and the normal state
properties.

\section{Experimental data and discussion}
The resistance and Hall effect measurements were done in a physical
property measurement system (Quantum Design, PPMS) with magnetic
field up to 9$\;$T. The six-lead method was used in the measurement
on the longitudinal and the transverse resistivity at the same time.
The resistance was measured by either sweeping magnetic field at a
fixed temperature or sweeping temperature at a fixed field. The
temperature stabilization was better than $0.1\%$ and the resolution
of the voltmeter was better than 10$\;$nV.

\subsection{Resistive transition and upper critical field}

\begin{figure}
\includegraphics[width=8cm]{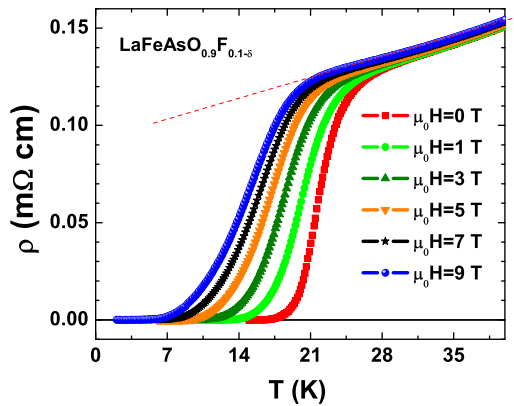}
\caption{(Color online) Temperature dependence of resistivity for
LaFeAsO$_{0.9}$F$_{0.1-\delta}$ bulk sample under different magnetic
fields. The onset transition point defined by $99\%\rho_\mathrm{n}$
shifts with the magnetic field weakly. The dashed line indicates the
extrapolation of the normal state resistivity. } \label{fig2}
\end{figure}

\begin{figure}
\includegraphics[width=8cm]{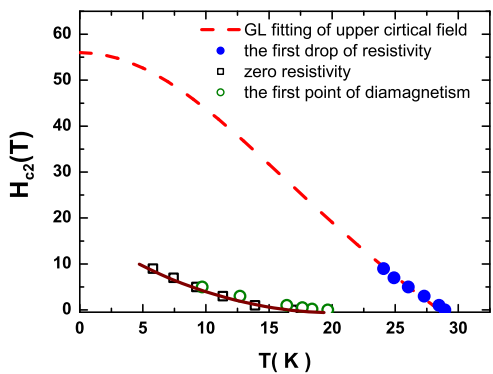}
\caption{(Color online) Phase diagram derived from the resistive
transition curves. The onset transition point gives rise to the
upper critical field $H_{\mathrm{c}2}^{ab}$ shown by the open
circles. The red solid line shows the theoretical curve based on the
GL theory (Eq.~\ref{eq2}). The magnetic onset transition point and
the zero resistivity point are quite close to each other, which are
shown by the open and filled squares, respectively. } \label{fig3}
\end{figure}

In Figure.~\ref{fig2} we present the temperature dependence of
resistivity for the LaFeAsO$_{0.9}$F$_{0.1-\delta}$ sample under
different magnetic fields. One can see that the onset transition
point shifts with the magnetic field weakly, but the zero resistance
point shifts more quickly to lower temperatures. This is
understandable since the latter is determined by the weak links
between the grains as well as the vortex flow behavior, while the
former is controlled by the upper critical field of the individual
grains. This allows to determine the upper critical field of this
material. However, we should mention that the upper critical field
determined in this way reflects mainly the situation of $H\parallel
ab$-plane since the Cooper pairs within the grains with this
configuration will last to the highest field compared to other
grains. Taking the onset point of the transition as the upper
critical field point $T_\mathrm{c}(H_{\mathrm{c}2})$ means that
almost all Cooper pairs are broken at this temperature and magnetic
field. By taking a criterion of $99\%\rho_\mathrm{n}$ (referenced by
the normal state resistivity $\rho_\mathrm{n}$ marked by the dashed
line in Figure.~\ref{fig3}) we can determine the upper critical
field $H_{\mathrm{c}2}(T)$ and shown in Figure.~\ref{fig3}. It is
seen that the slope of $H_{\mathrm{c}2}(T)$ near $T_\mathrm{c}$,
i.e., $\mathrm{d}H_{\mathrm{c}2}/\mathrm{d}T$ is about $-2.3\;$T/K.
By using the Werthamer-Helfand-Hohenberg (WHH) formula\cite{WHH} the
value of zero temperature upper critical field
$H_{\mathrm{c}2}^{ab}(0)$ can be estimated through:
\begin{equation}
H_{\mathrm{c}2}(0)=-0.693T_\mathrm{c}(\frac{\mathrm{d}H_{\mathrm{c}2}}{\mathrm{d}T})_{T=T_\mathrm{c}}
\label{eq1}
\end{equation}
Taking $T_\mathrm{c}= 28.9\;$K, we get $H_{\mathrm{c}2}(0) \approx
45.8 T$ roughly. We can also determine the $H_{\mathrm{c}2}$ by
using the formula based on the Ginzburg-Landau (GL) equation. In the
GL theory, it is known that $H_{\mathrm{c}2}=\Phi_0/2\pi\xi^2$ and
$\xi\propto \sqrt{(1+t^2)/(1-t^2)}$, with $\Phi_0$ the flux quanta,
$\xi$ the coherence length, $t=T/T_\mathrm{c}$ the reduced
temperature, thus one has
\begin{equation}
H_{\mathrm{c}2}(T)=H_{\mathrm{c}2}(0)\frac{1-t^2}{1+t^2} \label{eq2}
\end{equation}
Although the GL theory is specially applicable near $T_\mathrm{c}$,
above equation has been proved to be satisfied in much wider
temperature regime\cite{FangL}. We thus use above equation to fit
our data and show them as the solid line in Figure.~\ref{fig3}. The
zero temperature upper critical fields $H_{\mathrm{c}2}(0)$
determined in this way is $H_{\mathrm{c}2}(0)= 56\;$T. This value is
a bit higher than that obtained in using the WHH formula. We note
that a recent result reported that the WHH approximation could not
be simply applied in this material\cite{Fuchs}, and even the
$H_{\mathrm{c}2}(0)$ is affected by the multiband
property\cite{Hunte}. However our result can give a rough magnitude
of $H_{\mathrm{c}2}(0)$ because of the limit of magnetic field.

\subsection{Magnetoresistance}

\begin{figure}
\includegraphics[width=8cm]{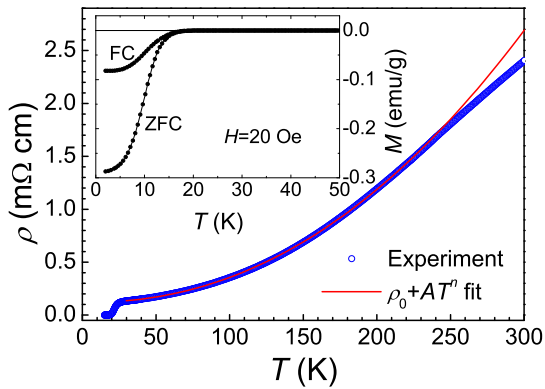}
\caption{(Color online) Temperature dependence of resistivity for
LaFeAsO$_{0.9}$F$_{0.1-\delta}$ bulk sample (shown by symbols), and
the solid line shows the fit in the low temperature range by
$\rho_0+AT^n$ with $\rho_0=0.117\;\mathrm{m}\Omega\,$cm and
$n=2.16$. Above about 220$\;$K, the fitting curve deviates clealy
from the data. The inset shows the temperature dependence of DC
magnetization for the zero field cooled (ZFC) and field cooled (FC)
at $H=20\;$Oe.} \label{fig4}
\end{figure}

In Figure.~\ref{fig4} we show the resistance transition of the
sample at zero field in wide temperature region by open circles.
From that we can get the zero resistance temperature at about
19$\;$K, the onset temperature 28.9$\;$K (99\% of the normal state
resistivity). The resistivity at 30$\;$K is 0.133$\;m\Omega\,$cm,
while the residual resistance ratio
$RRR\equiv\rho(300\;\mathrm{K})/\rho(30\;\mathrm{K})$ is about 18.1.
This may indicate a good quality of the sample in our experiments.
The general shape of the resistivity curve at this doping shows a
very good metallic behavior. The inset shown in Figure.~\ref{fig4}
gives the zero field cooled and also the field cooled DC
magnetization of the sample at 20$\;$Oe. The onset critical
temperature by magnetic measurement is about 24$\;$K (see by an
enlarged view), which is corresponding to the middle transition
point of resistance.

\begin{figure}
\includegraphics[width=8cm]{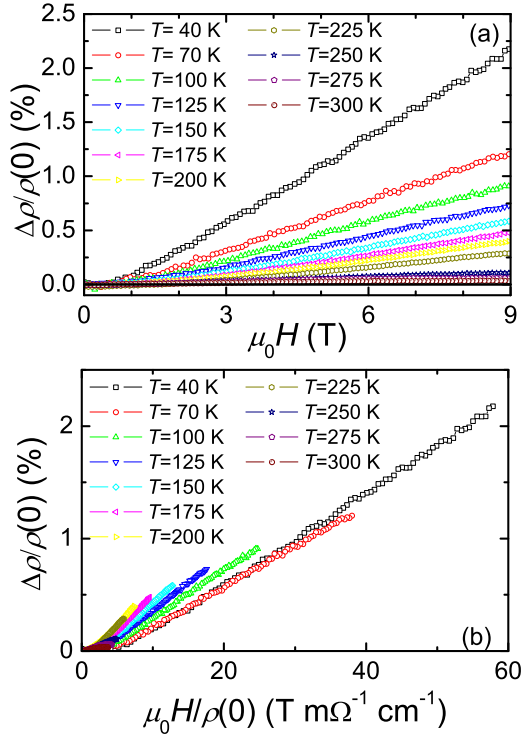}
\caption {(Color online) (a) Field dependence of megnetoresistance
$\Delta\rho/\rho(0)$ at different temperatures. One can see that the
MR decreases rapidly when the temperature is increased. Above about
250$\;$K, the MR becomes very weak. (b) Kohler plot for the sample
at different temperatures, and Kohler's rule is not obeyed.}
\label{fig5}
\end{figure}

Magnetoresistance is a very powerful tool to investigate the
electronic scattering process and the information about the Fermi
surface. For example, in MgB$_2$, a large magnetoresistance (MR) was
found which is closely related to the multiband
property.\cite{LiQ,YangHall} For this new superconductor, we also
measured and found a clear MR. In Figure.~\ref{fig5} (a) we present
field dependence of the MR ratio, i.e., $\Delta\rho/\rho_0$, where
$\rho$ is the resistivity, $\rho_0$ is the resistivity at zero
field, and $\Delta\rho=\rho(H)-\rho(0)$. One can see that the MR is
about 2.2\% at 40$\;$K and 9$\;$T for this sample. This ratio is one
order of magnitude smaller than MgB$_2$ with the same $RRR$.
However, considering that the sample is a polycrystalline sample,
the MR effect may be weakened by mixing the transport components
with the magnetic field along different directions of the
crystallographic axes. For a single band metal with a symmetric
Fermi surface, Kohler's law is normally obeyed. Kohler's
law\cite{Kohler} shows that the magnetoresistance
$\Delta\rho/\rho_0$ measured at different temperatures should be
scalable with the variable $H/\rho_0$. For MgB$_2$, Kohler'slaw is
not obeyed because of the multiband property.\cite{LiQ} We also do
scaling based on Kohler'slaw for this sample, the result is shown in
Figure.~\ref{fig5} (b). Clearly, the data measured at different
temperatures do not overlap and Kohler'slaw is violated for this
material. This discrepancy may suggest that in this new material
there may be multiband effect or other exotic scattering, such as
spin scattering which depends on the magnetic field. However, the MR
in this sample do show some specialty. For two-band or multiband
materials with weak MR, the MR ratios could be well described by the
expression $\Delta\rho/\rho_0\propto H^2$ in the low field region.
This is because the contribution of the higher order even terms of
$\mu_0H$ could be omitted at a low field (the odd terms are absent
here according to the Boltzmann equation for electronic transport).
This effect is also found in NbSe$_2$ which has a complex Fermi
surface structure.\cite{NbSe2} It was reported that both
LaFeAsO\cite{Singh} and LaFePO\cite{LaOFeP} have five orbitals
crossing the Fermi level, and Fermi surfaces with both electron and
hole type band are present. In this sense, the anomalies mentioned
above in the present new superconductor may also be induced by
multiband effect together with complex Fermi surfaces. On single
crystal samples, when the field is along $c$-axis, a stronger
in-plane magnetoresistance is expected.

\subsection{Hall effect}
\begin{figure}
\includegraphics[width=8cm]{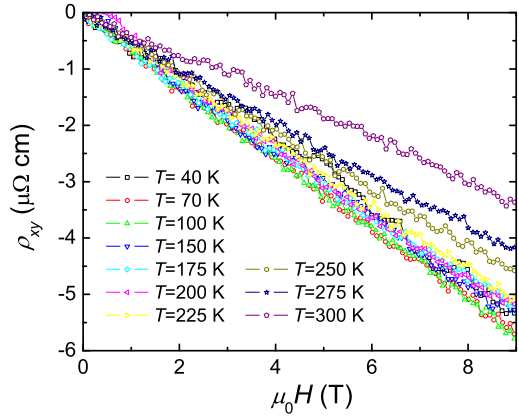}
\caption{(Color online) $\rho_{xy}$ versus the magnetic field
$\mu_0H$ at different temperatures. The curves at the temperatures
below 250$\;$K have similar behaviors, while at temperatures above
that temperature, the absolute values of slopes have a sudden
decrease.} \label{fig6}
\end{figure}

\begin{figure}
\includegraphics[width=8cm]{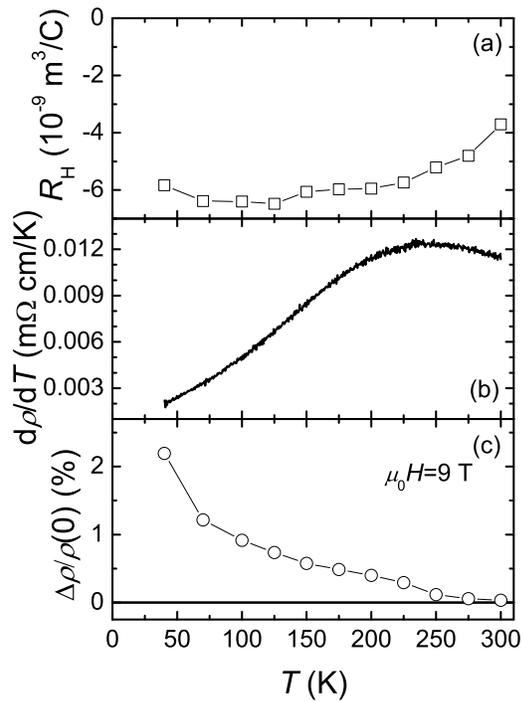}
\caption{Temperature dependence of the Hall coefficient (a),
differential of the $\rho-T$ curve (b), and the magnetoresistance
ratio at 9$\;$T (c). There are obvious turning point at about
230$\;$ for both of the curves shown in (a) and (b), while the
magnetoresistance shows a sudden decrease in that point.}
\label{fig7}
\end{figure}

For a normal metal with Fermi liquid feature, the Hall coefficient
is a constant versus temperature. However, this situation is changed
for a multiband material or a sample with non-Fermi liquid behavior,
such as the cuprate superconductors. We also do the Hall effect
measurement for this sample. As shown in Figure.~\ref{fig6}, the
transverse resistivity remains negative at all temperatures above
the critical temperature, indicating that the electric transport is
dominated by electron-like charge carriers, not hole-like ones. For
a clean sample, the nonlinear Hall effect is also a sign of
multiband, and the effect is weaker in dirtier samples.
\cite{YangHall} In our measurements, all curves shown in
Figure.~\ref{fig6} have good linearity versus the magnetic field
which may be caused by the disorders within the sample. From this
set of data, the Hall coefficient $R_\mathrm{H}=\rho_{xy}/H$ is
determined and shown in Figure.~\ref{fig7} (a). It is clear that
$R_\mathrm{H}$ is almost a constant below about 250$\;$K, and there
is  a kink at about 150$\;$K considering the small error showing in
the figure. The charge carrier density calculated by
$n=1/R_\mathrm{H} e$ is about $9.8\times10^{20}\;\mathrm{cm}^{-3}$
which is very close to the cuprate superconductors. This may imply
that the superfluid density in this superconductor is also very
diluted, as suggested by a recent theoretical proposal.\cite{Singh}
However, it should be noted that the Hall coefficient $R_\mathrm{H}$
could not be simply expressed as $1/ne$ for a multiband material,
\cite{YangHall} so it needs further consideration if multiband
property dominates the electric transport. From Figure.~\ref{fig6},
the transverse resistivity curves almost overlap at $T<250\;$K.
However at the temperature above 250$\;$K, the absolute value of the
slope, i.e., the Hall coefficient $R_\mathrm{H}$ decreases quickly.
As shown in Figure.~\ref{fig7}(a), $R_\mathrm{H}$ behaves a little
different from another recent work on
La(O$_{0.9}$F$_{1-\delta}$)FeAs.\cite{NLWang} In our sample,
$R_\mathrm{H}$ shows a weak $T$ dependence at the temperature below
225$\;$K; while at $T>250\;$K, the absolute value of $R_\mathrm{H}$
has a rapid decrease. This is similar to the situation of
MgCNi$_3$\cite{MgCNi}, which may be caused by the exotic scattering
when a ferromagnetic fluctuation is present. The temperature
dependent Hall coefficient also tells us that either the multiband
effect or some unusual scattering process may be involved in the
electron conduction in the material.

In order to reveal an anomalous behavior at about 230$\;$K, here we
make a further discussion. The differential of the $\rho-T$ curve
shown in Figure.~\ref{fig4} is given in Figure.~\ref{fig7} (b). It
is clear that there is a maximum value at about 240$\;$K of the
differential slope $\mathrm{d}\rho/\mathrm{d}T$. This is similar to
the situation reported in the underdoped LaOFeAs sample,
\cite{FirstJACS} where they found a clear kink point of resistivity
and marked the temperature point as $T_\mathrm{anom}$. In our
present sample, this kink is absent, but the resistivity shows a
weak downward feature around 240$\;$K, as evidenced by a bump on the
curve of $\mathrm{d}\rho/\mathrm{d}T$ vs. $T$. Therefore the anomaly
at about 240$\;$K in our sample may be related to the situation at
$T_\mathrm{anom}$ in the original paper\cite{FirstJACS}. We also try
to fit the normal state $\rho-T$ curve with the formula
$\rho_0+AT^n$. As shown in Figure.~\ref{fig4}, the curve below
240$\;$K could be well fitted and the fitting results give values of
$n=2.16$ and $\rho_0=0.117\;m\Omega\,$cm. However, just from this
temperature and above, the fitting curve starts deviating from the
data. As shown in Figure.~\ref{fig7} (c), there is also a sudden
decrease of MR at the temperature of 240$\;$K. The origin of this
anomaly at about 230$\;$K certainly needs further consideration.

So far, a weak magnetic signal has been measured in the sample in
the normal state, which is exactly the same as that reported in the
original paper by Kamihara et al.\cite{FirstJACS}. Magnetic
measurements show that this magnetic signal even has a small
hysteresis in low field region. At this moment, we don't know
whether this magnetic signal is due to an intrinsic feature of the
LaFeAsO phase, or is due to the second tiny impure phase. If the
former is right, some exotic scattering, like the electron-magnon
scattering or magnetic skew scattering would exist. This deserves a
further study on improved samples. Our present investigation will
thus provide a basic platform for the future studies.

\section{Conclusions}

In summary, the temperature dependence of resistivity under
different magnetic fields, the magnetoresistance and Hall
coefficient have been measured in the newly found layered
superconductor LaFeAsO$_{0.9}$F$_{0.1-\delta}$. The value of
$H_{\mathrm{c}2}$ at zero temperature is obtained roughly to be
about 50 Tesla. The Hall coefficient is negative indicating that the
electron-like charge carriers dominate the electrical transport. The
charge carrier density at 100$\;$K is about $9.8
\times10^{20}\;\mathrm{cm}^{-3}$ manifesting that the superconductor
may have a diluted superfluid, as in cuprate superconductors. The
Hall coefficient $R_\mathrm{H}$ has a weak temperature dependence
below 230$\;$K,  but it rises more rapidly above that temperature.
At the similar temperature the magnetoresistance becomes very small,
together with a maximum of the differential of the resistivity curve
$\rho(T)$. The $\rho(T)$ curve at low temperature could be fitted by
$\rho_0+AT^n$ with $n=2.16$. Kohler's law is clearly violated in all
temperature region. These observations can be explained by the
multi-band effect or some exotic scattering, like the scattering
with magnetic moments or in presence of weak magnetic correlation.

\section{Acknowledgements}
This work is supported by the Natural Science Foundation of China,
the Ministry of Science and Technology of China (973 project:
2006CB01000, 2006CB921802), the Knowledge Innovation Project of
Chinese Academy of Sciences (ITSNEM).


\section*{References}
\begin{thebibliography}{10}

\bibitem{FirstJACS}Kamihara Y, Watanabe T, Hirano M, and
Hosono H, 2008 {\it J. Am. Chem. Soc.} {\bf130} 3296

\bibitem{HHwen}Wen H H, Mu G, Fang L, Yang H, and Zhu X Y, 2008 {\it Europhys Lett.} {\bf82} 17009

\bibitem{xhchen}Chen X H, Wu T, Wu G, Liu R H, Chen H, and Fang  D F,
2008 {\it Nature.} {\bf453} 761

\bibitem{gfchen}Chen G F, Li Z, Wu D, Li G, Hu W Z , Dong J, Zheng  P, Luo J
L, and Wang N L, 2008 {\it Chin. Phys. Lett.} {\bf25} 2235

\bibitem{zhianren}Ren Z A, Yang J, Lu W, Yi W, Shen X L, Li Z C,
 Che G C, Dong X L, Sun L L, Zhou F and Zhao Z X, 2008 {\it EuroPhys. Lett.} {\bf82} 57002

\bibitem{zhanren}Ren Z A, Yang J, Lu W, et al., 2008 {\it Chin. Phys. Lett.} {\bf25}
2215

\bibitem{chengpeng} Chen P, Fang L, Yang H, Zhu X Y,
Mu G, Luo H Q, Wang Z S, and Wen H H,
 2008 {\it Science. in. China.} {\bf51} 719

\bibitem{WHH} Werthamer N R, Helfand E, Hohenberg P C, 1966 {\it Phys. Rev.} {\bf147} 295

\bibitem{FangL}Fang L, Wang Y, Zou P Y, Tang L, Xu Z, Chen H,
Dong C, Shan L and Wen H H,  2005 {\it Phys. Rev. B.} {\bf 72}
014534

\bibitem{Fuchs}Fuchs G, Drechsler S L, Kozlova N, Behr G, Koehler A, Werner J,
Nenkov, Hess C, Klingeler R, Hamann-Borrero J E , Kondrat A,
Grobosch M, Knupfer M, Freudenberger J, Buechner B, Schultz L, 2008
 {\it preprint} arXiv:0806.0781.


\bibitem{Hunte}Hunte F, Jaroszynski J, Gurevich A, Larbalestier D C,
 Jin R, Sefat A.S, McGuire M.A, Sales B.C, Christen D.K, and Mandrus D, 2008 {\it Nature} {\bf453},
 903.

\bibitem{LiQ}Qi Li, Liu B T, Hu Y F, Chen J, Gao H, Shan L, Wen H H,
Pogrebnyakov A V, Redwing J M, and Xi X X, 2006 {\it Phys. Rev.
Lett.} {\bf96} 167003

\bibitem{YangHall}Yang H, Liu Y, Zhuang C G, Shi J R, Yao Y G,
 Massidda S, Monni M, Jia Y, Xi X X, Li Q, Liu Z K, Feng Q R, 2008
 {\it preprint} arXiv:0802.3443

\bibitem{Kohler}Zinman J M, 2001 in {\it Electrons and Phonons} Classics
Series (Oxford University Press., New York,)


\bibitem{NbSe2}Corcoran R, Meeson P, Onuki Y, Probst P A,
Springford M, Takita K, Harima H, Guo G Y, and Gyorffy B L, 1994
{\it J. Phys.: Condens. Matter} {\bf6} 4479


\bibitem{Singh}Singh D J and Du M H, 2008 {\it preprint}
arXiv:0803.0429

\bibitem{LaOFeP}Lebgue S, 2007 {\it Phys. Rev. B.} {\bf75} 035110

\bibitem{NLWang}Chen G F, Li Z, Li G, Zhou J, Wu D, Dong J, Hu W Z, Zheng P,
 Chen Z J, Luo J L, and Wang N L, 2008 {\it preprint}
arXiv:0803.0128

\bibitem{MgCNi}Li S Y, Fan R, Chen X H, Wang C H, Mo W Q, Ruan K Q, Xiong Y M, Luo X G, Zhang H T,
Li L, Sun Z, and Cao L Z, 2001 {\it Phys. Rev. B.} {\bf64} 132505

\end{thebibliography}
\end{document}